\documentclass[prd,showpacs,preprintnumbers,amsmath,amssymb,nofootinbib]{revtex4}
\usepackage{cancel}
\usepackage{booktabs}
\usepackage{multirow}
\usepackage{slashed}
\usepackage{xcolor}
\usepackage{graphicx}
\usepackage[colorlinks,
            citecolor=blue,
            anchorcolor=red,
            menucolor=red,
            linkcolor=red,
            filecolor=red,
            runcolor=red,
            urlcolor=blue,
            frenchlinks=red]{hyperref}
\begin{document}
\title{Heavy-flavored tetraquark states with the $QQ\bar{Q}\bar{Q}$ configuration}

\author{Jing Wu$^1$, Yan-Rui Liu$^1$}
\email{yrliu@sdu.edu.cn} \affiliation{ $^1$School of Physics and Key
Laboratory of Particle Physics and Particle Irradiation (MOE),
Shandong University, Jinan 250100, China}
\author{Kan Chen$^{2,3}$, Xiang Liu$^{2,3}$}
\email{xiangliu@lzu.edu.cn} \affiliation{
$^2$School of Physical Science and Technology, Lanzhou University, Lanzhou 730000, China\\
$^3$Research Center for Hadron and CSR Physics, Lanzhou University
and Institute of Modern Physics of CAS, Lanzhou 730000, China }
\author{Shi-Lin Zhu$^{4,5,6}$}
\email{zhusl@pku.edu.cn} \affiliation{ $^4$School of Physics and
State Key Laboratory of Nuclear Physics and Technology, Peking
University, Beijing 100871, China
\\
$^5$Collaborative Innovation Center of Quantum Matter, Beijing
100871, China
\\
$^6$Center of High Energy Physics, Peking University, Beijing
100871, China }

\begin{abstract}

In the framework of the color-magnetic interaction, we
systematically investigate the mass spectrum of the tetraquark
states composed of four heavy quarks with the $QQ\bar Q\bar Q$
configuration in this work. We also show their strong decay
patterns. Stable or narrow states in the $bb\bar{b}\bar{c}$ and $bc\bar{b}\bar{c}$ systems are found to be possible. We hope the studies shall be helpful to the experimental search for
heavy-full exotic tetraquark states.

\end{abstract}

\maketitle

\section{Introduction}\label{sec1}

Searching for exotic states is an interesting research field full of challenges and opportunities. Among the various exotic candidates,
the multiquark state is a very popular hadronic configuration. In fact, the concept of the multiquark state was proposed long time ago
in Refs. \cite{Jaffe:1976ig,Jaffe:1976ih,Jaffe:1976yi}. In the past decades, experimentalists have been trying to find them. The
observations of the charmonium-like/bottomonium-like $XYZ$ states \cite{Swanson:2006st,Zhu:2007wz,Voloshin:2007dx,Drenska:2010kg,Chen:2016qju,Chen:2016heh,Hosaka:2016pey,Ali:2017jda}
have provided valuable hints of the existence of the exotic states. Especially, these charged charmonium-like or bottomonium-like states
like the $Z(4430)$ \cite{Choi:2007wga,Mizuk:2009da,Chilikin:2014bkk,Aaij:2014jqa}, the $Z_1(4050)$ \cite{Mizuk:2008me}, the $Z_2(4250)$
\cite{Mizuk:2008me}, the $Z_c(3900)$ \cite{Ablikim:2013mio,Liu:2013dau,Xiao:2013iha,Ablikim:2015tbp}, the
$Z_c(3885)$ \cite{Ablikim:2013xfr,Ablikim:2015swa,Ablikim:2015gda}, the $Z_c(4020)$ \cite{Ablikim:2013wzq,Ablikim:2014dxl}, the
$Z_c(4025)$ \cite{Ablikim:2013emm,Ablikim:2015vvn}, the $Z_c(4200)$ \cite{Chilikin:2014bkk}, the $Z_b(10610)$ \cite{Belle:2011aa}, and
the $Z_b(10650)$ \cite{Belle:2011aa} have forced us to consider the existence of the multiquark matter very seriously. So many
observations in the heavy quark sector in recent years are surprising, which provides new opportunities for us to understand
the nature of strong interaction.

Experimentalists continue to surprise us after the observations of $XYZ$ states. In 2015, the LHCb Collaboration reported the
hidden-charm pentaquarks $P_c(4380)$ and $P_c(4450)$ in the invariant mass of $\psi p$\footnote{For
convenience, here and in the following, $\psi$ means $J/\psi$ once the symbol is adopted.}. They are consistent with the pentaquark
structures \cite{Wu:2010jy,Wu:2010vk,Yang:2011wz,Wang:2011rga,Yuan:2012wz,Wu:2012md,Garcia-Recio:2013gaa,Xiao:2013yca,Uchino:2015uha}.
With a model-independent analysis, the LHCb confirms their previous model-dependent evidence for these states \cite{Aaij:2016phn}. Very
recently, a structure $X(5568)$ was announced by the D\O\, Collaboration \cite{D0:2016mwd}. This narrow state is about 200 MeV
below the $B\bar{K}$ threshold and decays into $B_s^0\pi^\pm$. Therefore, the $X(5568)$ may be a typical tetraquark state composed
of four different flavors \cite{Agaev:2016mjb,Wang:2016mee,Wang:2016tsi,Chen:2016mqt,Zanetti:2016wjn,Agaev:2016ijz,Liu:2016ogz,Agaev:2016lkl,Dias:2016dme,Wang:2016wkj,He:2016yhd,
Stancu:2016sfd,Tang:2016pcf} while the molecule state assignment to $X(5568)$ is not favored \cite{Agaev:2016urs}. However, the LHCb
Collaboration did not confirm the existence of the $X(5568)$ \cite{LHCb:2016ppf}, which makes some theorists consider the
difficulty of explaining the $X(5568)$ as a genuine resonance \cite{Burns:2016gvy,Guo:2016nhb,Lu:2016zhe,Esposito:2016itg,Albaladejo:2016eps,Ali:2016gdg,Albuquerque:2016nlw,Chen:2016npt,Jin:2016cpv}.
Although there exist different opinions of the $X(5568)$, the observation of the $X(5568)$ again ignites theorist's enthusiasm of
exploring exotic tetraquark states.

As discussed above, there exist possible candidates of the hidden-charm tetraquark states and the tetraquark with a heavy
flavor quark and three light quarks. If the multiquark states indeed exist in nature, we have a strong reason to believe that there are
more tetraquark states with other flavor configurations. In this work, we focus on the heavy-flavor tetraquark states with the
$Q_1Q_2\bar{Q}_3\bar{Q}_4$ configuration, where $Q_i$ is a $c$ or $b$ quark. Although these tetraquark states are still missing in
experiment, it is time to carry out a systematic investigation of the mass spectrum of the $Q_1Q_2\bar{Q}_3\bar{Q}_4$ tetraquark
system, which may provide important information to further experimental exploration.

In Ref. \cite{Iwasaki:1975pv}, Iwasaki studied the hidden-charm tetraqark state composed of a pair of charmed quark and anti-quark as well as a pair of light quark and anti-quark, which has the $c\bar{c}q\bar{q}$ configuration. The dimeson configuration ($Q^2\bar{q}^2$) is stable against strong decay into two mesons \cite{Ballot:1983iv}. Chao performed a systematical investigation of the $cc\bar{c}\bar{c}$ tetraquark system in a quark-gluon model for
the first time in Ref. \cite{Chao:1980dv}. In Ref. \cite{Heller:1985cb}, the authors used the Born-Oppenheimer
approximation for heavy quarks in the MIT bag and found that the heavy-quark system $c^2\bar{c}^2$ is stable against breakup into two $c\bar{c}$ pairs. But in a potential model calculation \cite{Ader:1981db}, the authors suggested that for identical quarks, there is no stable $QQ\bar{Q}\bar{Q}$ state. Similar opinions were shared by the authors of Ref. \cite{SilvestreBrac:1993ss}. However, in Ref. \cite{Lloyd:2003yc}, Lloyd and Vary obtained bound tetraquark states by adopting a parameterized Hamiltonian to compute the spectrum of the $cc\bar{c}\bar{c}$ tetraquark states. The tetraquark spectrum was also studied with a generalization of the hyperspherical harmonic formalism in Ref. \cite{Barnea:2006sd}. In addition, the calculation of the chromomagnetic interaction for the $Q^2\bar{Q}^2$ system was performed in Ref. \cite{SilvestreBrac:1992mv}. In understanding the nature of the $Y(4260)$ meson in Ref. \cite{Chiu:2005ey}, as a byproduct, a lattice study indicates that the $J^{PC}=1^{--}$ $cc\bar{c}\bar{c}$ state is possible. If the P-wave $cc\bar{c}\bar{c}$ state exists where the orbital angular momentum contributes some repulsion, the lower ground tetraquark states should also exist. Recently, there are further discussions about the properties of fully-heavy tetraquarks \cite{Hughes:2017xie,Bai:2016int, Anwar:2017toa, Wang:2017jtz, Debastiani:2017msn, Karliner:2016zzc, Eichten:2017ual, Bicudo:2017usw,Vega-Morales:2017pmm,Richard:2017vry,Chen:2016jxd}.

In this work, we calculate the mass splittings of the $Q_1Q_2\bar{Q}_3\bar{Q}_4$ tetraquark system in a simple quark model
systematically. A typical feature for such tetraquarks is that the isospin is always zero and the flavor wave function is always
symmetric for identical quarks. One may only focus on the color-spin part when the Pauli principle is employed to exclude some
configurations. For the interaction between the heavy quarks, the short-range gluon exchange force is a dominant source. In the
one-gluon-exchange potential model, the color-spin or color-magnetic interaction part is responsible for the mass splitting of the ground
hadrons with the same flavor content. In the present study, we will adopt the color-magnetic interaction (CMI) to perform the
calculations, which violates the heavy quark symmetry.

We organize the paper as follows. After the introduction, we present
the formalism of calculation in Sec. \ref{sec2}. We give the
numerical results and discuss the possible decay modes in Sec.
\ref{sec3}. We summarize our results in the final section.

\section{Formalism}\label{sec2}

The color-magnetic interaction in the model reads
\begin{eqnarray}
H_{CM}=-\sum_{i,j}C_{ij}\lambda_{i}\cdot\lambda_{j}\sigma_{i}\cdot\sigma_{j},
\end{eqnarray}
where $\lambda_{i}$'s are the Gell-Mann matrices and $\sigma_i$'s are the Pauli matrices. The above Hamiltonian was deduced from the
one-gluon-exchange interaction \cite{DeRujula:1975qlm}. The effective coupling constants $C_{ij}$ incorporate effects from the
spatial wave function and the quark masses, which depend on the system. We determine their values in the next section. This
Hamiltonian leads to a mass formula for the studied system
\begin{eqnarray}\label{modelhamiltonian}
{M}=\sum_{i}m_{i}+\langle H_{CM}\rangle,
\end{eqnarray}
where $m_{i}$ is the effective mass of the $i$-th constituent quark, which includes the constituent quark mass and effects from other
terms such as color-electric interaction and color confinement.

To calculate the matrix elements for the color-spin interaction, one may construct the $color\otimes spin$ wave functions explicitly and
calculate them by definition. In studying multiquark systems, a simpler way was used in Refs.
\cite{Hogaasen:2004pm,Hogaasen:2005jv,Buccella:2006fn}. One just calculates the matrix elements in color space and in spin space
separately with the Hamiltonaians $H_{C}=-\sum_{i,j}C_{ij}\lambda_{i}\cdot\lambda_{j}$ and $H_{S}=-\sum_{i,j}C_{ij}\sigma_{i}\cdot\sigma_{j}$. Then $\langle
H_{CM}\rangle$ is obtained by combining the results of $\langle H_C\rangle$ and $\langle H_S\rangle$ with the common coefficient $C_{ij}$.

In order to consider the constraint from the Pauli principle, we use a diquark-antidiquark picture to analyze the configurations. Here the notation diquark just means two quarks. In the spin space, the allowed wave functions read
\begin{eqnarray}
\chi_{1}=|(Q_{1}Q_{2})_{1}(\bar{Q}_{3}\bar{Q}_{4})_{1}\rangle_{2},\quad
\chi_{2}=|(Q_{1}Q_{2})_{1}(\bar{Q}_{3}\bar{Q}_{4})_{1}\rangle_{1},\nonumber\\
\chi_{3}=|(Q_{1}Q_{2})_{1}(\bar{Q}_{3}\bar{Q}_{4})_{1}\rangle_{0},\quad
\chi_{4}=|(Q_{1}Q_{2})_{1}(\bar{Q}_{3}\bar{Q}_{4})_{0}\rangle_{1},\nonumber\\
\chi_{5}=|(Q_{1}Q_{2})_{0}(\bar{Q}_{3}\bar{Q}_{4})_{1}\rangle_{1},\quad
\chi_{6}=|(Q_{1}Q_{2})_{0}(\bar{Q}_{3}\bar{Q}_{4})_{0}\rangle_{0},
\end{eqnarray}
where the subscripts on the right hand side denote the spin of the $Q_1Q_2$, spin of the $\bar{Q}_3\bar{Q}_4$, and that of the system, respectively.
According to the $SU(3)$ group theory, the diquark in the color space belongs to the representation $6_c$ or $\bar{3}_c$, while the
anti-diquark's representation is $\bar{6}_c$ or $3_c$. Then one has two kinds of color-singlet state
\begin{eqnarray}
\phi_{1}=|(Q_{1}Q_{2})^{6}(\bar{Q}_{3}\bar{Q}_{4})^{\bar{6}}\rangle,\quad
\phi_{2}=|(Q_{1}Q_{2})^{\bar{3}}(\bar{Q}_{3}\bar{Q}_{4})^{3}\rangle.
\end{eqnarray}

Combine the spin and color wave functions, we get twelve possible $color\otimes spin$ wave functions for the
$Q_{1}Q_{2}\bar{Q}_{3}\bar{Q}_{4}$ system
\begin{eqnarray}\label{color-spin-wave}
&\phi_1\chi_1=|(Q_{1}Q_{2})^{6}_{1}(\bar{Q}_{3}\bar{Q}_{4})^{\bar{6}}_{1}\rangle_{2}\delta_{12}\delta_{34},\quad
&\phi_2\chi_1=|(Q_{1}Q_{2})^{\bar{3}}_{1}(\bar{Q}_{3}\bar{Q}_{4})^{3}_{1}\rangle_{2},\nonumber\\
&\phi_1\chi_2=|(Q_{1}Q_{2})^{6}_{1}(\bar{Q}_{3}\bar{Q}_{4})^{\bar{6}}_{1}\rangle_{1}\delta_{12}\delta_{34},\quad
&\phi_2\chi_2=|(Q_{1}Q_{2})^{\bar{3}}_{1}(\bar{Q}_{3}\bar{Q}_{4})^{3}_{1}\rangle_{1},\nonumber\\
&\phi_1\chi_3=|(Q_{1}Q_{2})^{6}_{1}(\bar{Q}_{3}\bar{Q}_{4})^{\bar{6}}_{1}\rangle_{0}\delta_{12}\delta_{34},\quad
&\phi_2\chi_3=|(Q_{1}Q_{2})^{\bar{3}}_{1}(\bar{Q}_{3}\bar{Q}_{4})^{3}_{1}\rangle_{0},\nonumber\\
&\phi_1\chi_4=|(Q_{1}Q_{2})^{6}_{1}(\bar{Q}_{3}\bar{Q}_{4})^{\bar{6}}_{0}\rangle_{1}\delta_{12},\quad
&\phi_2\chi_4=|(Q_{1}Q_{2})^{\bar{3}}_{1}(\bar{Q}_{3}\bar{Q}_{4})^{3}_{0}\rangle_{1}\delta_{34},\nonumber\\
&\phi_1\chi_5=|(Q_{1}Q_{2})^{6}_{0}(\bar{Q}_{3}\bar{Q}_{4})^{\bar{6}}_{1}\rangle_{1}\delta_{34},\quad
&\phi_2\chi_5=|(Q_{1}Q_{2})^{\bar{3}}_{0}(\bar{Q}_{3}\bar{Q}_{4})^{3}_{1}\rangle_{1}\delta_{12},\nonumber\\
&\phi_1\chi_6=|(Q_{1}Q_{2})^{6}_{0}(\bar{Q}_{3}\bar{Q}_{4})^{\bar{6}}_{0}\rangle_{0},\quad
&\phi_2\chi_6=|(Q_{1}Q_{2})^{\bar{3}}_{0}(\bar{Q}_{3}\bar{Q}_{4})^{3}_{0}\rangle_{0}\delta_{12}\delta_{34},
\end{eqnarray}
where the used notation is $|(Q_{1}Q_{2})^{color}_{spin}(\bar{Q}_{3}\bar{Q}_{4})^{color}_{spin}\rangle_{spin}$.
We insert a symbol $\delta_{ij}$ in the wave functions to reflect the constraint from the Pauli principle. When the $i$-th quark and
the $j$-th quark are the same, $\delta_{ij}=0$. Otherwise, $\delta_{ij}=1$.

Replacing $Q_i$ with $c$ or $b$ quark, one gets nine possibilities for the flavor content, six of which need to be studied:
$bb\bar{b}\bar{b}$, $cc\bar{c}\bar{c}$, $bb\bar{c}\bar{c}$, $bb\bar{b}\bar{c}$, $cc\bar{c}\bar{b}$, and $bc\bar{b}\bar{c}$. The
other three cases $cc\bar{b}\bar{b}$, $cb\bar{b}\bar{b}$, and $bc\bar{c}\bar{c}$ correspond to the antiparticles of
$bb\bar{c}\bar{c}$, $bb\bar{b}\bar{c}$, and $cc\bar{c}\bar{b}$, respectively. So their formulas are not independent. Here,
considering the Pauli principle, one may categorize the six systems into three sets: (1) $bb\bar{b}\bar{b}$, $cc\bar{c}\bar{c}$, and
$bb\bar{c}\bar{c}$, where $\delta_{12}=\delta_{34}=0$; (2) $bb\bar{b}\bar{c}$ and $cc\bar{b}\bar{c}$, where $\delta_{12}=0,\delta_{34}=1$; and (3) $bc\bar{b}\bar{c}$ where $\delta_{12}=\delta_{34}=1$. The number of independent wave functions for them is 4, 6, and 12, respectively. We present the
resulting CMI matrix elements system by system. By the way, the usually mentioned ``good'' diquark (spin=$0$, color=$\bar{3}_c$) exists only in the systems
containing the $(bc)$ substructure since the most attractive $(bb)^{\bar{3}_c}_{spin\text{-}0}$ and $(cc)^{\bar{3}_c}_{spin\text{-}0}$ objects are forbidden by the Pauli principle.

\subsection{The $bb\bar{b}\bar{b}$ and $cc\bar{c}\bar{c}$ systems}
The color-spin structures of these two systems are the same. The
only difference lies in the quark mass. So we put them together for
discussions.

The $bb\bar{b}\bar{b}$ system is a neutral state. Its possible
quantum numbers are $I^G(J^{PC})=0^+(2^{++})$, $0^-(1^{+-})$, or
$0^+(0^{++})$. The number of states is constrained by the Pauli
principle. For the case $J=2$, the wave function is
$\phi_{2}\chi_{1}$ and the obtained $\langle H_{CM}\rangle$ is given
by $\frac{16}{3}(C_{bb}+C_{b\bar{b}})$. The color-magnetic
interaction is certainly repulsive. For the case $J=1$, the wave
function is $\phi_{2}\chi_{2}$ and $\langle
H_{CM}\rangle=\frac{16}{3}(C_{bb}-C_{b\bar{b}})$. Since the
quark-quark interaction and the quark-antiquark interaction is
related with the G-parity transformation, the CMI for this $J=1$
state is expected to be weak. For the case $J=0$, there are two
possible wave functions $\phi_{2}\chi_{3}$ and $\phi_{1}\chi_{6}$.
Although the color-spin structures are different, their mixing is
allowed by the color-magnetic interaction. The obtained symmetric
CMI matrix in the $(\phi_{2}\chi_{3},\phi_{1}\chi_{6})^T$ base is
\begin{eqnarray}
\langle H_{CM}\rangle= \left(\begin{array}{cc}
\frac{16}{3}(C_{bb}-2C_{b\bar{b}}) & 8\sqrt{6}C_{b\bar{b}} \\
                                   & 8C_{bb}
\end{array}\right).
\end{eqnarray}

The $cc\bar{c}\bar{c}$ system has similar expressions. For the case
$J=2$, $\langle H_{CM}\rangle=\frac{16}{3}(C_{cc}+C_{c\bar{c}})$
with the color-spin wave function $\phi_{2}\chi_{1}$. For the case
$J=1$, $\langle H_{CM}\rangle=\frac{16}{3}(C_{cc}-C_{c\bar{c}})$
with the wave function $\phi_{2}\chi_{2}$. For $J=0$, the CMI matrix
is
\begin{eqnarray}
\langle H_{CM}\rangle= \left(\begin{array}{cc}
\frac{16}{3}(C_{cc}-2C_{c\bar{c}}) & 8\sqrt{6}C_{c\bar{c}} \\
                               & 8C_{cc}
\end{array}\right).
\end{eqnarray}

The above compact tetraquark system has the same quark content as
the molecular states composed of two bottomonium (charmonium)
mesons. However, the interaction between heavy quarks is dominantly
through the short-range gluon exchange. Once the interaction is
strong enough to bind the two mesons, the resulting object very
probably tends to form a compact structure instead of a
loosely-bound molecule.

In the extreme case that the molecules exist, the configuration
mixing is possible. For the $S$-wave $\Upsilon\Upsilon$ ($\psi\psi$)
state, the allowed quantum numbers are $I^G(J^{PC})=0^+(2^{++})$ or
$0^+(0^{++})$ since the wave function in spin space should be
symmetric for identical mesons. The quantum numbers for the $S$-wave
state composed of two $\eta_b$ ($\eta_c$) mesons are only
$0^+(0^{++})$. For the state composed of one $\eta_b$ ($\eta_c$) and
one $\Upsilon$ ($\psi$), the quantum numbers are just
$I^G(J^{PC})=0^-(1^{+-})$. Therefore, the allowed $J^{PC}$ appear
both in the molecule and diquark-antidiquark pictures. The number of
states is also equivalent. Considering that the interaction between
heavy quarks is through the gluon exchange force, one does not
expect large mass difference between these two configurations. The
compact structure may contribute significantly to the properties of
the molecules with the same quantum numbers.

If the tetraquark states have larger masses, the quark
rearrangements into the meson-meson channels with the same quantum
numbers may happen. We will discuss such possible decay patterns
after their masses are estimated.

\subsection{The $bb\bar{c}\bar{c}$ and $cc\bar{b}\bar{b}$ systems}

The two systems are related through the $C$-parity transformation
and they have the same color-spin matrix elements. The possible
quantum numbers of them are $I(J^P)=0(2^+)$, $0(1^+)$, or $0(0^+)$.
Again the Pauli principle results in four states for the
$bb\bar{c}\bar{c}$ (or $cc\bar{b}\bar{b}$) system. The color-spin
wave functions are $\phi_{2}\chi_{1}$ and $\phi_{2}\chi_{2}$ for the
case of $J=2$ and $J=1$, respectively. The corresponding $\langle
H_{CM}\rangle$'s are $\frac{8}{3}(C_{bb}+C_{cc}+2C_{b\bar{c}})$ and
$\frac{8}{3}(C_{bb}+C_{cc}-2C_{b\bar{c}})$. For the case $J=0$, the
allowed wave functions are the same as those of the previous
systems, $\phi_{2}\chi_{3}$ and $\phi_{1}\chi_{6}$. Consider their
mixing, one gets

\begin{eqnarray}
\langle H_{CM}\rangle=\left(\begin{array}{cc}
\frac{8}{3}(C_{bb}+C_{cc}-4C_{b\bar{c}}) & 8\sqrt{6}C_{b\bar{c}} \\
                               & 4(C_{bb}+C_{cc}) \\
\end{array}\right).
\end{eqnarray}

These systems have the same quark content as the $B_c^{(*)}B_c^{(*)}$ meson-meson states. The quantum numbers of $B_c^{*-}B_c^{*-}$ ($B_c^{*+}B_c^{*+}$) are $I(J^P)=0(2^+)$ or $0(0^+)$, those of $B_c^-B_c^-$ ($B_c^+B_c^+$) are $0(0^+)$. There is only one $B_c^-B_c^{*-}$ ($B_c^+B_c^{*+}$) state with $I(J^P)=0(1^+)$.

\subsection{The $bb\bar{b}\bar{c}$ and $cb\bar{b}\bar{b}$ systems}

The two systems have the same matrix elements. Their quantum numbers
are also $I(J^P)=0(2^+)$, $0(1^+)$, or $0(0^+)$. We here focus on
the $bb\bar{b}\bar{c}$ system. Now the number of the vector states
is 3 and that of the scalar states is 2. For the $J=2$ case, the
color-spin wave function is again $\phi_2\chi_1$. The resulting
color-magnetic matrix element is $\langle
H_{CM}\rangle=\frac{8}{3}(C_{bb}+C_{bc}+C_{b\bar{b}}+C_{b\bar{c}})$.
For the case $J=1$, three possible wave functions $\phi_2\chi_2$,
$\phi_2\chi_4$, and $\phi_1\chi_5$ are allowed. The last one has
different color structure from the other two. Although their mixing
occurs, from the obtained matrix [base: $(\phi_2\chi_2,
\phi_2\chi_4, \phi_1\chi_5)^T$]
\begin{eqnarray}
\langle H_{CM}\rangle= \left(\begin{array}{ccc}
\frac{8}{3}(C_{bb}+C_{bc}-C_{b\bar{b}}-C_{b\bar{c}})
   &\frac{8\sqrt{2}}{3}(C_{b\bar{b}}-C_{b\bar{c}})&8(C_{b\bar{b}}-C_{b\bar{c}})\\
   &\frac{8}{3}(C_{bb}-3C_{bc})&-4\sqrt{2}(C_{b\bar{c}}+C_{b\bar{b}}) \\
   &  & \frac{4}{3}(3C_{bb}-C_{bc})
\end{array}\right),
\end{eqnarray}
one observes that the mixing strength for $\phi_2\chi_2$ and
$\phi_2\chi_4$ and that for $\phi_2\chi_2$ and $\phi_1\chi_5$ may be
both small. The remaining case is for $J=0$, where possible wave
functions are $\phi_2\chi_3$ and $\phi_1\chi_6$. Now, one has
\begin{eqnarray}
\langle H_{CM}\rangle= \left(\begin{array}{cc}
\frac{8}{3}(C_{bb}+C_{bc}-2C_{b\bar{b}}-2C_{b\bar{c}})&4\sqrt{6}(C_{b\bar{b}}+C_{b\bar{c}})\\
                               & 4(C_{bb}+C_{bc})
\end{array}\right).
\end{eqnarray}

The meson-meson systems with the quark content $bb\bar{b}\bar{c}$
are $\Upsilon B_c^-$, $\Upsilon B_c^{*-}$, $\eta_bB_c^-$, and
$\eta_bB_c^{*-}$. Their quantum numbers are $I(J^P)=0(1^+)$,
$0([2,1,0]^+)$, $0(0^+)$, and $0(1^+)$, respectively.

\subsection{The $cc\bar{c}\bar{b}$ and $bc\bar{c}\bar{c}$ systems}

The situation is similar to the $bb\bar{b}\bar{c}$ and
$cb\bar{b}\bar{b}$ systems. By exchanging $b$ and $c$ there, one
easily gets relevant matrix elements. For comparison, we focus on
the $cc\bar{c}\bar{b}$ system. For the case $J=2$, one has $\langle
H_{CM}\rangle=\frac{8}{3}(C_{cc}+C_{bc}+C_{b\bar{c}}+C_{c\bar{c}})$.
For the case $J=1$, the matrix for the color-spin interaction reads
\begin{eqnarray}
\langle H_{CM}\rangle= \left(\begin{array}{ccc}
\frac{8}{3}(C_{cc}+C_{bc}-C_{b\bar{c}}-C_{c\bar{c}})
       &\frac{8\sqrt{2}}{3}(C_{c\bar{c}}-C_{b\bar{c}})&8(C_{c\bar{c}}-C_{b\bar{c}})\\
       &\frac{8}{3}(C_{cc}-3C_{bc})&-4\sqrt{2}(C_{c\bar{c}}+C_{b\bar{c}})\\
       &  & \frac{4}{3}(3C_{cc}-C_{bc})
\end{array}\right).
\end{eqnarray}
For the case $J=0$, the matrix is
\begin{eqnarray}
\langle H_{CM}\rangle =\left(\begin{array}{cc}
\frac{8}{3}(C_{cc}+C_{bc}-2C_{b\bar{c}}-2C_{c\bar{c}}) & 4\sqrt{6}(C_{b\bar{c}}+C_{c\bar{c}}) \\
                               & 4(C_{cc}+C_{bc})
\end{array}\right).
\end{eqnarray}

The signs for the non-diagonal matrix elements seem to be
inconsistent with the previous systems after the replacements $b\to
c$ and $c\to b$. Actually they do not affect the final results. For
detailed argument, one may consult Eq. (2) of Ref. \cite{Liu:2011xc}
and relevant explanations there.

The meson-meson states that these tetraquarks can rearrange into are
$\psi B_c$, $\psi B_c^*$, $\eta_c B_c$, and $\eta_c B_c^*$.

\subsection{The $bc\bar{b}\bar{c}$ system}

The Pauli principle does not give any constraints on the wave
functions now. The two wave functions for $J=2$ and the four wave
functions for $J=0$ will mix, respectively. However, one should be
careful in discussing the mixing with the six wave functions for
$J=1$ because the system is neutral and may have $C$-parity.

If $J=2$, both the diquark and the antidiquark have angular momentum
1. The state should have definite $C$-parity + and the quantum
numbers are $I^G(J^{PC})=0^+(2^{++})$. With the base
$(\phi_1\chi_1,\phi_2\chi_1)^T$, one may get the CMI matrix
\begin{eqnarray}
\langle H_{CM}\rangle= \left(\begin{array}{cc}
-\frac{2}{3}(4C_{bc}-5C_{b\bar{b}}-10C_{b\bar{c}}-5C_{c\bar{c}})
     &2\sqrt{2}(C_{b\bar{b}}-2C_{b\bar{c}}+C_{c\bar{c}})\\
     &\frac{4}{3}(4C_{bc}+C_{b\bar{b}}+2C_{b\bar{c}}+C_{c\bar{c}})
\end{array}\right).
\end{eqnarray}

If $J=0$, both the diquark and the antidiquark have the same angular
momentum. The quantum numbers for the system are
$I^G(J^{PC})=0^+(0^{++})$. The obtained CMI matrix is
\begin{eqnarray}
&&\langle H_{CM}\rangle=\nonumber\\
&&\left(\begin{array}{cccc}
-\frac{4}{3}\left(\begin{array}{c}2C_{bc}+5C_{b\bar{b}}\\+10C_{b\bar{c}}+5C_{c\bar{c}}\end{array}\right)
   & -\frac{10}{\sqrt{3}}(C_{b\bar{b}}-2C_{b\bar{c}}+C_{c\bar{c}}) & 4\sqrt{2}(C_{b\bar{b}}-2C_{b\bar{c}}+C_{c\bar{c}}) & 2\sqrt{6}(C_{b\bar{b}}+2C_{b\bar{c}}+C_{c\bar{c}}) \\
   & 8C_{bc} & 2\sqrt{6}(C_{b\bar{b}}+2C_{b\bar{c}}+C_{c\bar{c}}) & 0 \\
   &   & \frac{8}{3}(2C_{bc}-C_{b\bar{b}}-2C_{b\bar{c}}-C_{c\bar{c}}) & -\frac{4}{\sqrt{3}}(C_{b\bar{b}}-2C_{b\bar{c}}+C_{c\bar{c}}) \\
   &   &   & -16C_{bc}
\end{array}\right),\nonumber\\
\end{eqnarray}
where the base is
$(\phi_1\chi_3,\phi_1\chi_6,\phi_2\chi_3,\phi_2\chi_6)^T$.

If $J=1$, the states $\phi_1\chi_2$ and $\phi_2\chi_2$ have negative
$C$-parities. All the other four wave functions $\phi_1\chi_4$,
$\phi_2\chi_4$, $\phi_1\chi_5$, and $\phi_2\chi_5$ are not invariant
under $C$-parity transformation. But we may construct four states
which are invariant under $C$-parity transformations. The basic
procedure is similar to that given in Ref. \cite{Liu:2013rxa}.
Explicitly, the two $C=+$ states are
\begin{eqnarray}
{[\phi\chi]}_+^{6\bar{6}}&=&\frac{1}{\sqrt2}(\phi_1\chi_4+\phi_1\chi_5),\nonumber\\
{[\phi\chi]}_+^{\bar{3}3}&=&\frac{1}{\sqrt2}(\phi_2\chi_4+\phi_2\chi_5),
\end{eqnarray}
and the two $C=-$ states are
\begin{eqnarray}
{[\phi\chi]}_-^{6\bar{6}}&=&\frac{1}{\sqrt2}(\phi_1\chi_4-\phi_1\chi_5),\nonumber\\
{[\phi\chi]}_-^{\bar{3}3}&=&\frac{1}{\sqrt2}(\phi_2\chi_4-\phi_2\chi_5).
\end{eqnarray}
Only states with the same quantum numbers may mix. So we have two
color-spin matrices. For the states with $I^G(J^{PC})=0^+(1^{++})$,
the matrix is
\begin{eqnarray}
\langle H_{CM}\rangle= \left(\begin{array}{cc}
\frac23(4C_{bc}+5C_{b\bar{b}}+5C_{c\bar{c}}-10C_{b\bar{c}})&-2\sqrt2(C_{b\bar{b}}+C_{c\bar{c}}+2C_{b\bar{c}})\\
&\frac43(-4C_{bc}+C_{b\bar{b}}+C_{c\bar{c}}-2C_{b\bar{c}})
\end{array}\right),
\end{eqnarray}
with the base
$({[\phi\chi]}_+^{6\bar{6}},{[\phi\chi]}_+^{\bar{3}3})^T$. For the
states with $I^G(J^{PC})=0^-(1^{+-})$, the matrix reads
\begin{eqnarray}
\langle H_{CM}\rangle= \left(\begin{array}{cccc}
-\frac23\left(\begin{array}{c}4C_{bc}+5C_{b\bar{b}}\\+5C_{c\bar{c}}+10C_{b\bar{c}}
\end{array}\right)&2\sqrt2(C_{b\bar{b}}+C_{c\bar{c}}-2C_{b\bar{c}})&\frac{20}{3}(C_{b\bar{b}}-C_{c\bar{c}})&-4\sqrt2(C_{b\bar{b}}-C_{c\bar{c}})\\
&\frac43\left(\begin{array}{c}4C_{bc}-C_{b\bar{b}}\\-C_{c\bar{c}}-2C_{b\bar{c}}
\end{array}\right)&-4\sqrt2(C_{b\bar{b}}-C_{c\bar{c}})&\frac83(C_{b\bar{b}}-C_{c\bar{c}})\\
&&\frac23\left(\begin{array}{c}4C_{bc}-5C_{b\bar{b}}\\-5C_{c\bar{c}}+10C_{b\bar{c}}
\end{array}\right)&2\sqrt2(C_{b\bar{b}}+C_{c\bar{c}}+2C_{b\bar{c}})\\
&&&-\frac43\left(\begin{array}{c}4C_{bc}+C_{b\bar{b}}\\+C_{c\bar{c}}-2C_{b\bar{c}}\end{array}\right)
\end{array}\right),\nonumber\\
\end{eqnarray}
where the base is $(\phi_1\chi_2, \phi_2\chi_2,
{[\phi\chi]}_-^{6\bar{6}},{[\phi\chi]}_-^{\bar{3}3})^T$.

There are two kinds of molecule configurations for the
$bc\bar{b}\bar{c}$ system. In the bottomonium+charmonium case, the
allowed quantum numbers are $I^G(J^{PC})=0^+(0^{++})$ for the
$\eta_b\eta_c$ system, $0^-(1^{+-})$ for the $\eta_b\psi$ system,
$0^-(1^{+-})$ for the $\Upsilon\eta_c$, and $0^+([2,1,0]^{++})$ for
the $\Upsilon\psi$ system. In the meson-antimeson case, those for
$B_c^-B_c^+$ are $0^+(0^{++})$, those for $B_c^-B_c^{*+}\pm
B_c^{*-}B_c^{+}$ are $0^{\pm}(1^{+\pm})$, and those for
$B_c^{*-}B_c^{*+}$ are $0^+([2,0]^{++})$ or $0^-(1^{+-})$.

\section{Numerical results}\label{sec3}

\subsection{Parameters}

We need to determine six coefficients $C_{b\bar{b}}$,
$C_{c\bar{c}}$, $C_{b\bar{c}}$, $C_{bb}$, $C_{cc}$, and $C_{bc}$ in
discussing the mass splittings for various
$Q_1Q_2\bar{Q}_3\bar{Q}_4$ systems. Their masses may be further
estimated with the Hamiltonian in Eq. (\ref{modelhamiltonian}) once
the effective quark masses $m_c$ and $m_b$ are used.

From the mass splitting between $\Upsilon$ and $\eta_b$,
$m_\Upsilon-m_{\eta_b}=[\frac{16}{3}C_{b\bar{b}}]-[-16C_{b\bar{b}}]=61$
MeV \cite{pdg}, one extracts $C_{b\bar{b}}=2.9$ MeV. Similarly, the
value of $C_{c\bar{c}}=5.3$ MeV is obtained from the mass splitting
$m_{J/\psi}-m_{\eta_c}=114$ MeV. Since the excited $B_c^*$ meson has
not been observed yet, we just estimate the value of $C_{b\bar{c}}$
to be 3.3 MeV from $m_{B_c^*}-m_{B_c}=70$ MeV calculated with a
quark model \cite{Godfrey:1985xj}. Although the $\Xi_{cc}^{++}$ baryon was confirmed recently by the LHCb Collaboration \cite{Aaij:2017ueg} after the first observation at SELEX \cite{Mattson:2002vu,Ocherashvili:2004hi}, the available heavy baryon masses are still not enough for us to extract the value of $C_{cc}$. Here we perform our calculation with the approximation $C_{QQ}=C_{Q\bar{Q}}$ ($Q=c$, $b$). Since there is no dynamics in the present model, the choice of the approximation to determine $C_{QQ}$ is not unique. However, the results induced by the change of $C_{QQ}$ should not be large \cite{Lee:2007tn,Lee:2009rt}. For comparison, we also adopt the approximation $\frac{C_{QQ}}{C_{Q\bar{Q}}}=\frac{C_{nn}}{C_{n\bar{n}}}\approx\frac23$ and check the extreme case $C_{QQ}=0$, where $C_{nn}=18.4$ MeV is extracted from the light baryon masses \cite{Wu:2016gas}. By using the mass difference between $\rho$ and $\pi$, one gets $C_{n\bar{n}}=29.8$ MeV. The latter approximation certainly gives a smaller $C_{QQ}$. If the interactions within the diquarks are effectively attractive (repulsive), the approximation $\frac{C_{QQ}}{C_{Q\bar{Q}}}\approx\frac23$ should result in heavier (lighter) tetraquark states. Here, the effective interaction within diquarks reflects the repulsion or attraction effect from the enhancement or cancellation between the quark-quark (and antiquark-antiquark) interaction in the case of channel coupling. For comparison, we use these two approximations in our estimation. The values of the quark-quark coupling parameters estimated with them are listed in Table \ref{parameter}.
\begin{table}[!h]
\caption{Two sets of effective coupling parameters in units of MeV with different approximations.}\label{parameter}
\centering
\begin{tabular}{ccc}
\hline\hline
&$\begin{array}{c}\text{Set I}\\ \left(C_{QQ}=C_{Q\bar{Q}}\right)\end{array}$& $\begin{array}{c}\text{Set II}\\ \left(\frac{C_{QQ}}{C_{Q\bar{Q}}}=\frac{C_{nn}}{C_{n\bar{n}}}\right)\end{array}$\\
\hline
$C_{cc}$&5.3&3.3\\
$C_{bb}$&2.9&1.9\\
$C_{bc}$&3.3&2.0\\\hline\hline
\end{tabular}
\end{table}

To determine the masses of the tetraquarks, we adopt two approaches
in the present work: 1) One estimates the meson masses with the
effective heavy quark masses $m_b=5052.9$ MeV and $m_c=1724.8$ MeV. These
values were adopted in understanding the strange properties of tetraquark states \cite{Wu:2016gas,Chen:2016ont} and the pentaquark states \cite{Wu:2017weo} ;
2) One calculates the masses from a meson-meson threshold, where the mass formula is $M=M_{th}-\langle H_{CM}\rangle_{th}+\langle H_{CM}\rangle$ and the
relevant meson masses are: $m_\Upsilon=9460.3$ MeV,
$m_{\psi}=3096.9$ MeV, and $m_{B_c}=6275.1$ MeV \cite{pdg}. The
latter method has been used in estimating the mass of an exotic
$T_{cc}$ \cite{Hyodo:2012pm,Hyodo:2017hue}.

\subsection{The $bb\bar{b}\bar{b}$ and $cc\bar{c}\bar{c}$ systems}

\begin{figure}[htbp]
\centering
\begin{tabular}{ccc}
\includegraphics[width=200pt]{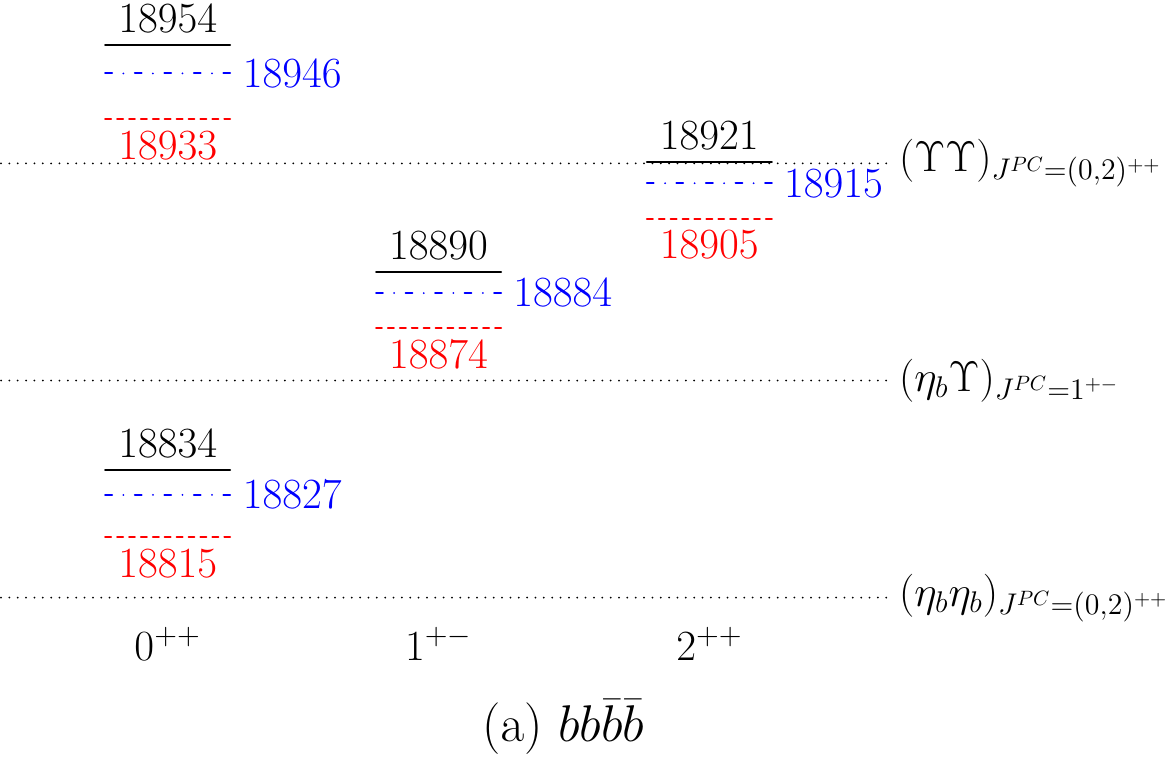}&\quad&\includegraphics[width=200pt]{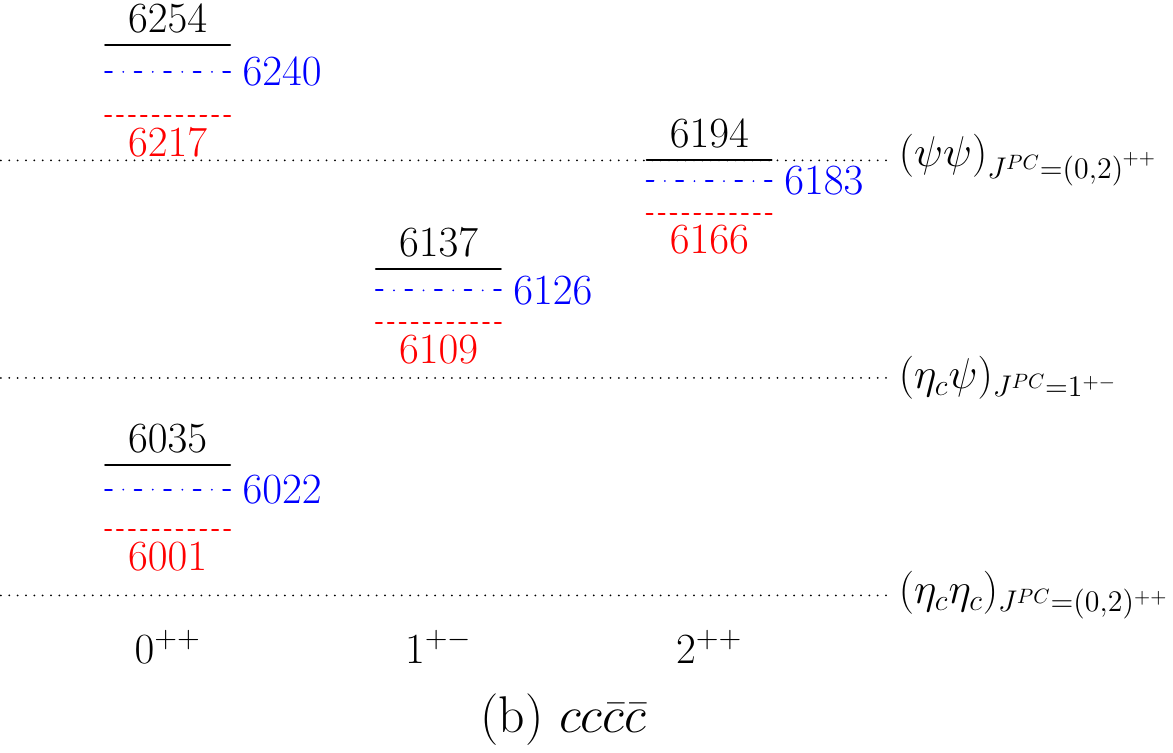}\label{cccc}\\
\\
\includegraphics[width=200pt]{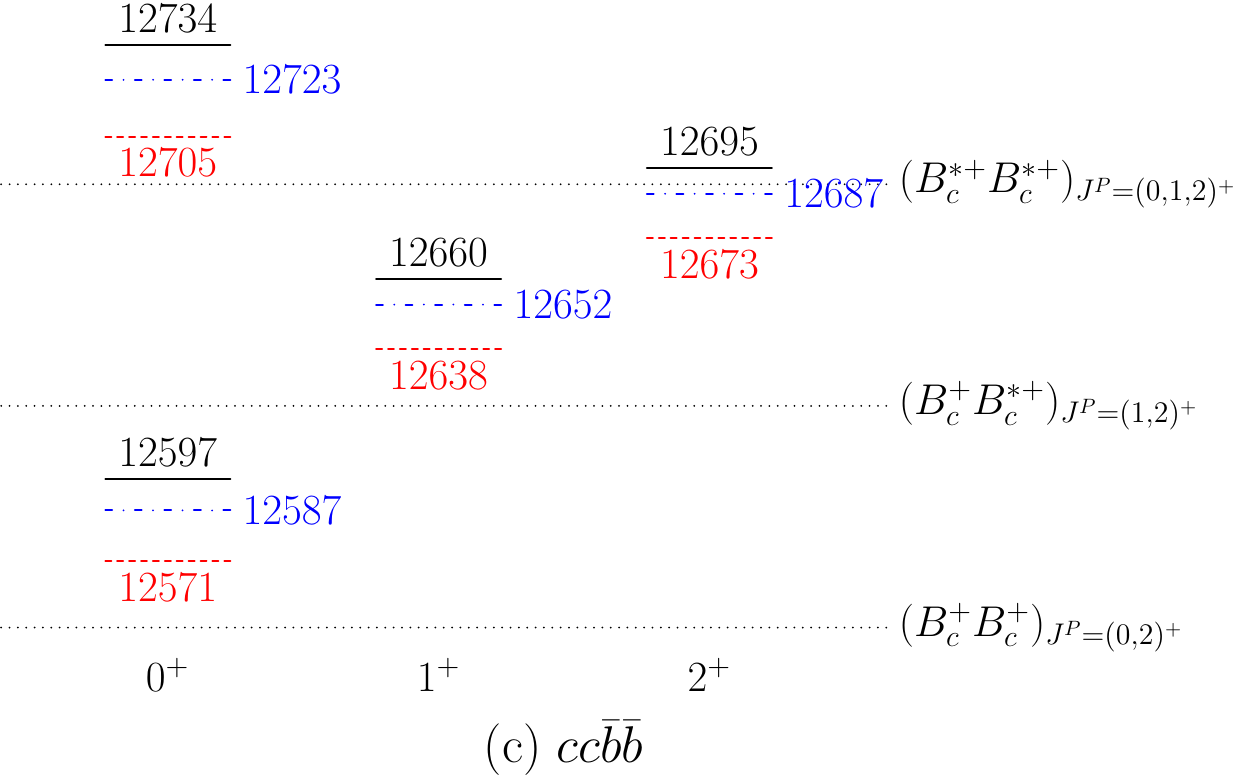}&\quad&\includegraphics[width=200pt]{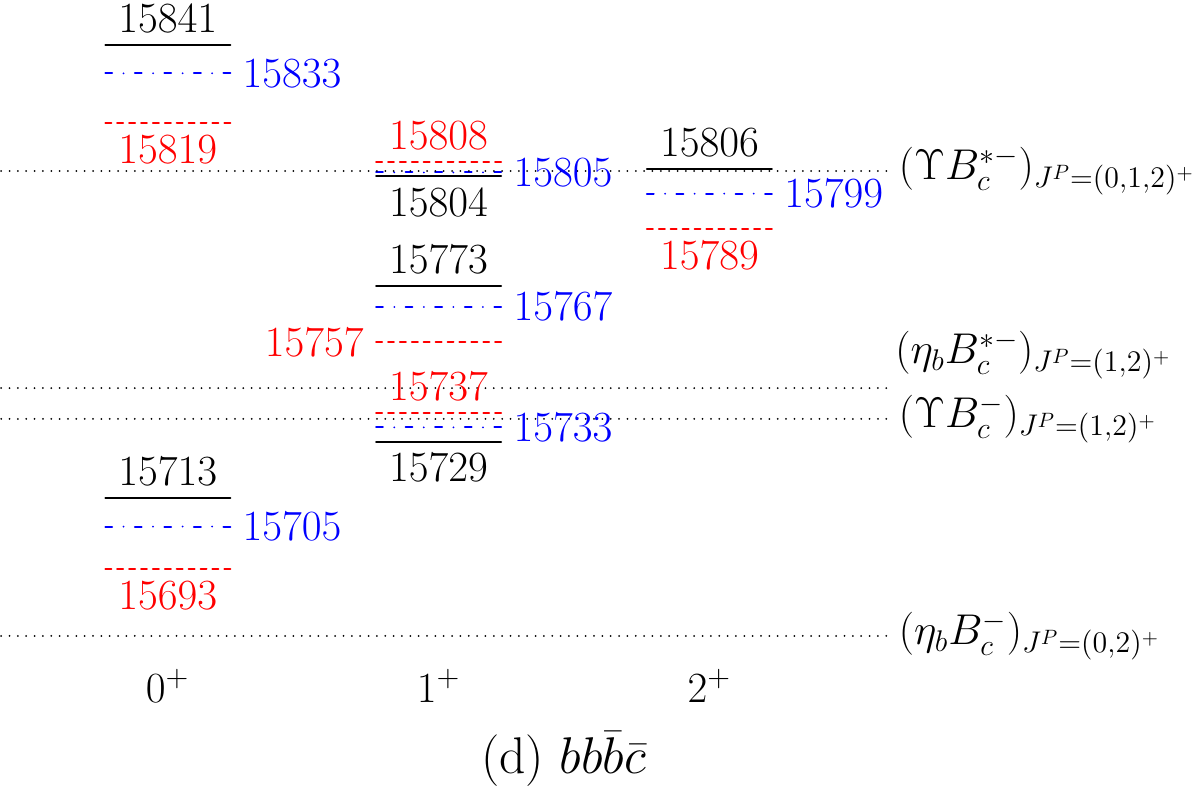}\\
\\
\includegraphics[width=200pt]{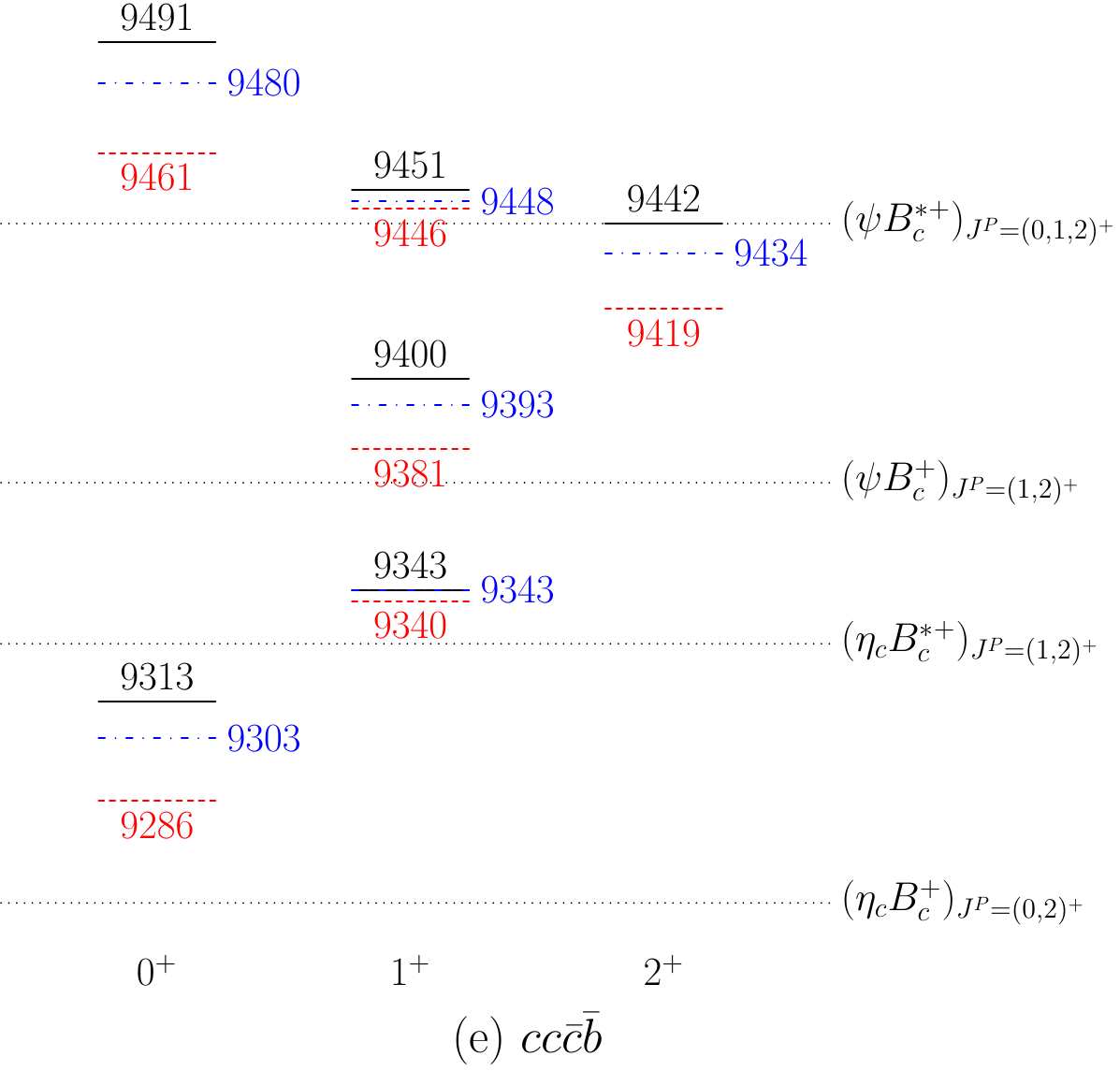}&\quad&\includegraphics[width=200pt]{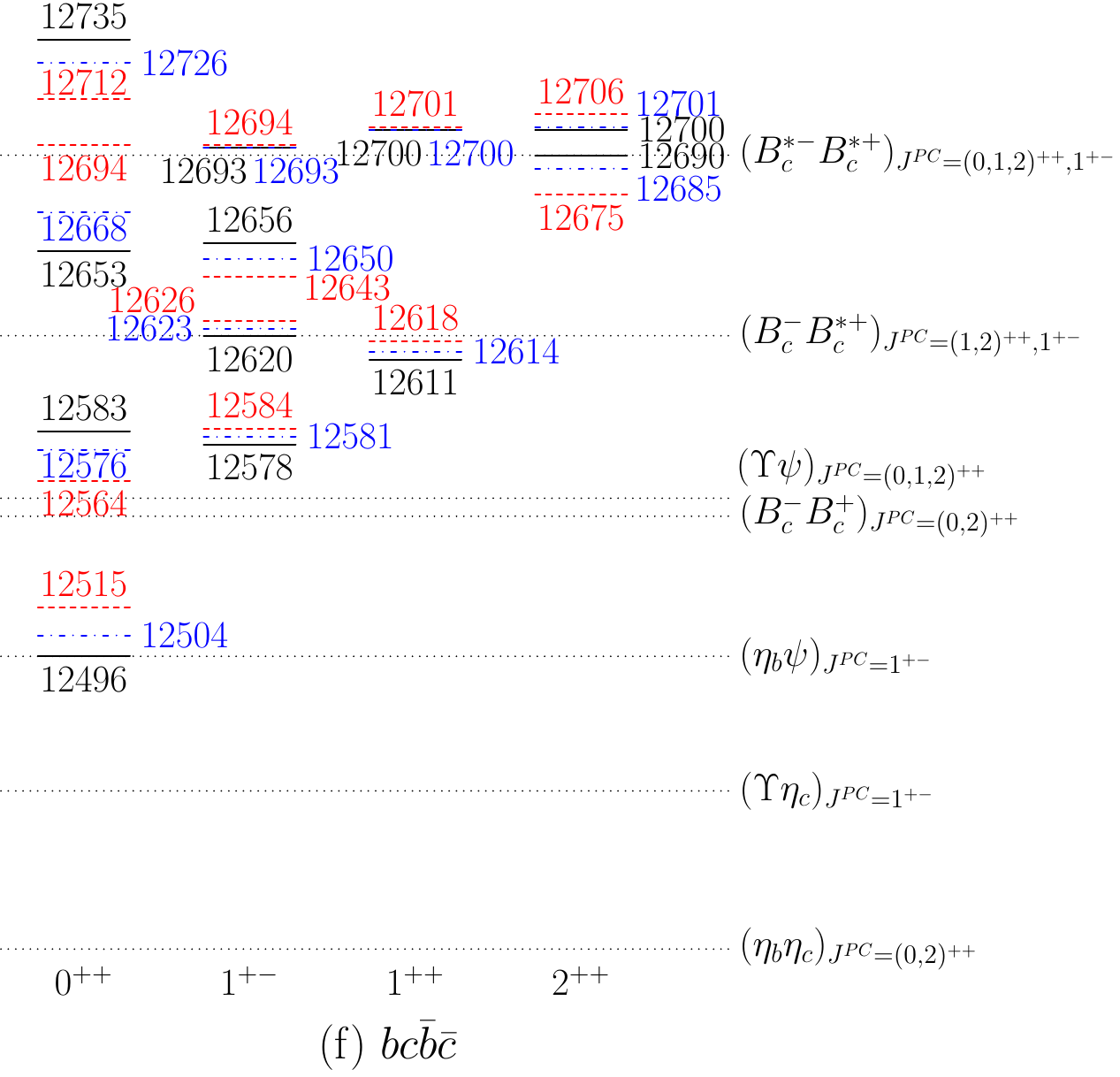}\\
\end{tabular}
\caption{Relative positions for the considered tetraquark states. The solid (black) and dashdotted (blue) lines correspond to masses estimated with the approximations $C_{QQ}=C_{Q\bar{Q}}$ and $C_{QQ}=\frac{C_{nn}}{C_{n\bar{n}}}C_{Q\bar{Q}}$, respectively. The dashed (red) lines are for the case $C_{QQ}=0$. The dotted lines indicate various meson-meson thresholds. When the quantum numbers in the subscript of the symbol for a meson-meson state are equal to the $J^{PC}$ (or $J^P$) of an initial state, the decay for the initial state into that meson-meson channel through S- or D-wave is allowed. We adopt the masses estimated with the reference thresholds of (a) $\Upsilon\Upsilon$, (b) $\psi\psi$, (c) $B_cB_c$, (d) $\Upsilon B_c$, (e) $\psi B_c$, and (f) $B_cB_c$. The masses are all in units of MeV.}\label{fig}
\end{figure}

It is easy to get the numerical results for the CMI matrix elements
with the above two sets of parameters. Adopting the approximation $C_{QQ}=C_{Q\bar{Q}}$, we obtain the CMI matrices, their eigenvalues and corresponding eigenvectors, and estimated masses with two different approaches. These results are presented in Table \ref{result:bbbb}. In the approximation $C_{QQ}=\frac{C_{nn}}{C_{n\bar{n}}}C_{Q\bar{Q}}$, the estimated masses of the tetraquark states are slightly different from those in the former approximation. In the following discussions, we mainly use the masses estimated with the parameters in Set I. Since the hadron masses estimated with the effective quarks are usually like an upper limit \cite{Wu:2016gas,Chen:2016ont,Wu:2017weo,Luo:2017eub}, we focus on the results estimated with reference thresholds. We assume that these masses are all reasonable values. To have an impression for the spectrum, we plot relative positions for the $bb\bar{b}\bar{b}$ ($cc\bar{c}\bar{c}$) states in Fig. \ref{fig} (a) [(b)]. The solid/black lines are for the approximation $C_{QQ}=C_{Q\bar{Q}}$ and the dashdotted/blue lines are for the approximation $C_{QQ}=\frac{C_{nn}}{C_{n\bar{n}}}C_{Q\bar{Q}}$. We also show the results in the extreme case $C_{QQ}=0$ with the dashed/red lines. The uncertainty caused by the change of $C_{QQ}$ is less than 20 (37) MeV in the $bb\bar{b}\bar{b}$ ($cc\bar{c}\bar{c}$) case.

\begin{table}[htbp]
\caption{Results for the $bb\bar{b}\bar{b}$ and $cc\bar{c}\bar{c}$
systems in units of MeV with the approximation $C_{QQ}=C_{Q\bar{Q}}$. The masses in the fifth column are
estimated with $m_b=5052.9$ MeV and $m_c=1724.8$ MeV. The last column
lists masses estimated from the $(\Upsilon\Upsilon)$ or $(J/\psi
J/\psi)$ threshold. The base for the $J=0$ case is
$(\phi_{2}\chi_{3},\phi_{1}\chi_{6})^T$.}\label{result:bbbb}
\centering
\begin{tabular}{c|cccccc}\hline
System& $J^{PC}$ & $\langle H_{CM} \rangle$ &Eigenvalue &Eigenvector &Mass&$(\Upsilon\Upsilon)/(\psi\psi)$\\
& $2^{++}$ &30.9&30.9&1&20243&18921\\
$(bb\bar{b}\bar{b})$&$1^{+-}$ &0.0&0.0&1&20212&18890\\
& $0^{++}$ &$\left(\begin{array}{cc}-15.5&56.8\\56.8&23.2\end{array}\right)$&$\left[\begin{array}{c}63.9\\-56.2\end{array}\right]$&$\left[\begin{array}{c}(0.58,0.81)\\(-0.81,0.58)\end{array}\right]$&$\left[\begin{array}{c}20275\\20155\end{array}\right]$&$\left[\begin{array}{c}18954\\18834\end{array}\right]$\\\hline\hline
& $2^{++}$ &56.5&56.5&1&6956&6194\\
$(cc\bar{c}\bar{c})$&$1^{+-}$ &0.0&0.0&1&6899&6137\\
& $0^{++}$ &$\left(\begin{array}{cc}-28.3&103.9\\103.9&42.4\end{array}\right)$&$\left[\begin{array}{c}116.8\\-102.6\end{array}\right]$&$\left[\begin{array}{c}(0.58,0.81)\\(-0.81,0.58)\end{array}\right]$&$\left[\begin{array}{c}7016\\6797\end{array}\right]$&$\left[\begin{array}{c}6254\\6035\end{array}\right]$\\
\hline
\end{tabular}
\end{table}

The mass splitting between the scalar tetraquarks is around 120 MeV for the bottom system and 220 MeV for the charmed system. One finds that the mixing between different color structures is important here, which enlarges the mass difference between these two states. If one does not consider the mixing, the masses for the bottom case are 18913 MeV and 18875 MeV. Both states are below the $\Upsilon\Upsilon$ threshold and above the $\eta_b\Upsilon$ threshold. Once the mixing is considered, the higher state ($6_c$ $bb$ dominates) becomes a state above the $\Upsilon\Upsilon$ threshold while the lower one ($\bar{3}_c$ $bb$ dominates) becomes a state below the $\eta_b\Upsilon$ threshold. Certainly the mass shift affects decay properties. The charmed case is similar.

From the diagrams (a) and (b) in Fig. \ref{fig}, the estimated tetraquark masses are all above the lowest meson-meson threshold. This observation is consistent with those in Refs. \cite{Karliner:2016zzc,Hughes:2017xie,Richard:2017vry}. The lowest $bb\bar{b}\bar{b}$ mass in \cite{Karliner:2016zzc} and ours are similar. From these diagrams, the masses obtained with parameters in Set II are all lower than those in Set I. This means that the interactions within the diquarks are effectively repulsive, which can be verified from the Hamiltonian expressions. If stable multiquark states need attractive diquarks, these $bb\bar{b}\bar{b}$ and $cc\bar{c}\bar{c}$ compact tetraquarks would tend to become meson-meson states because the quark-antiquark interaction is usually attractive (see the diagonal matrix elements in the Hamiltonian expressions) and these tetraquarks should not be stable.

If the studied states do exist, finding out their decay properties are helpful to the search at experiments. Possible rearrangement decay modes are easy to be understood from Fig. \ref{fig}. Since the feature for the $cc\bar{c}\bar{c}$ system is very similar to the bottom case, we here only concentrate on the latter one. For the $J=2$ tetraquark, the present model estimation gives a mass around the $\Upsilon\Upsilon$ threshold. If the approximation $C_{QQ}\approx \frac23C_{Q\bar{Q}}$ is more appropriate, the state is blow the threshold and it should have a relatively narrow width through $D$-wave decay into $\eta_b\eta_b$. It is a basic feature that high spin multiquark states have dominantly $D$-wave decay modes and should not be very broad \cite{Chen:2016ont,Wu:2017weo,Luo:2017eub}. For the $J=1$ tetraquark, its mass is 30 MeV above the $\eta_b\Upsilon$ threshold. From the quantum numbers, its rearrangement decay channel is only this $\eta_b\Upsilon$. For the $J=0$ tetraquarks, the higher one can decay into both $\Upsilon\Upsilon$ and $\eta_b\eta_b$ through both $S$- and $D$-wave interactions while the lower one decays only into $\eta_b\eta_b$. If we use their masses to denote these states, probably the ordering of the widths is $18954>18890\sim18834>18921$.

From Fig. \ref{fig} (a)-(b), the feature that all the states can decay is consistent with the feature that the effective interaction within the diquarks is repulsive. If we want to find relatively stable compact tetraquarks in Fig. \ref{fig}, good candidates should be those states for which dashed/red lines are above solid/black lines. Although the $bb\bar{b}\bar{b}$ and $cc\bar{c}\bar{c}$ systems do not satisfy this condition, we will see that such systems exist.

\subsection{The $bb\bar{c}\bar{c}$ and $cc\bar{b}\bar{b}$ systems}

\begin{table}[htbp]
\caption{Results for the $bb\bar{c}\bar{c}$ and $cc\bar{b}\bar{b}$
systems in units of MeV with the approximation $C_{QQ}=C_{Q\bar{Q}}$. The masses in the fifth column are
estimated with $m_b=5052.9$ MeV and $m_c=1724.8$ MeV. The last column
lists masses estimated from the $(B_cB_c)$ threshold. The base for
the $J=0$ case is
$(\phi_{2}\chi_{3},\phi_{1}\chi_{6})^T$.}\label{result:bbcbarcbar}
\centering
\begin{tabular}{c|ccccc}\hline
$J^P$ & $\langle H_{CM} \rangle$ &Eigenvalue &Eigenvector &Mass&$(B_cB_c)$\\
$2^+$ &39.5&39.5&1&13595&12695\\
$1^+$ &4.3&4.3&1&13560&12660\\
$0^+$ &$\left(\begin{array}{cc}-13.3&64.7\\64.7&32.8\end{array}\right)$&$\left[\begin{array}{c}78.4\\-58.9\end{array}\right]$&$\left[\begin{array}{c}(0.58,0.82)\\(-0.82,0.58)\end{array}\right]$&$\left[\begin{array}{c}13634\\13496\end{array}\right]$&$\left[\begin{array}{c}12734\\12597\end{array}\right]$\\
\hline
\end{tabular}
\end{table}

We present the numerical results for the $cc\bar{b}\bar{b}$ states in Table \ref{result:bbcbarcbar}. The $bb\bar{c}\bar{c}$ states are antiparticles of the $cc\bar{b}\bar{b}$ states and have the same results. Now the mass splitting between different spins is less than 140 MeV. This number lies between the splittings for the $bb\bar{b}\bar{b}$ case and the $cc\bar{c}\bar{c}$ case and thus the mixing effect is in the middle of the two. To understand the decay properties easily, we plot the spectrum for the $cc\bar{b}\bar{b}$ system and relevant meson-meson thresholds in Fig. \ref{fig} (c).

Basically, the behaviors for the rearrangement decays are similar to those for the $bb\bar{b}\bar{b}$ and $cc\bar{c}\bar{c}$ systems. The main difference lies in the $C$ parity. The former states have definite $C$ parities while the $cc\bar{b}\bar{b}$ states not. Without the condition of $C$ parity conservation, the $2^+$ $cc\bar{b}\bar{b}$ tetraquark can also decay into the $B_c^+B_c^{*+}$ channel through $D$-wave.

Up to now, experiments confirm only the ground $B_c$ meson. It means that the axial vector tetraquark decaying into $B_cB_c^*$ may not be observed in the near future. In the case that the $B_c^*$ meson is confirmed with enough data, the search for the $1^+$ $cc\bar{b}\bar{b}$ (or $bb\bar{c}\bar{c}$) tetraquark is also possible. However, the interactions within the diquarks are effectively repulsive and these tetraquarks should not be stable.

\subsection{The $bb\bar{b}\bar{c}$ and $cb\bar{b}\bar{b}$ systems}

\begin{table}[h!]
\caption{Results for the $bb\bar{b}\bar{c}$ and $cb\bar{b}\bar{b}$
systems in units of MeV with the approximation $C_{QQ}=C_{Q\bar{Q}}$. The masses in the fifth column are
estimated with $m_b=5052.9$ MeV and $m_c=1724.8$ MeV. The last column
lists masses estimated from the $(\Upsilon B_c)$ threshold. The base
for the $J=1$ case is $(\phi_2\chi_2, \phi_2\chi_4, \phi_1\chi_5)^T$
and that for the $J=0$ case is
$(\phi_{2}\chi_{3},\phi_{1}\chi_{6})^T$.}\label{result:bbbbarcbar}
\centering
\begin{tabular}{c|ccccc}\hline
$J^P$ & $\langle H_{CM} \rangle$ &Eigenvalue &Eigenvector &Mass&$(\Upsilon B_c)$\\
$2^+$ &33.1&33.1&1&16917&15806\\
$1^+$ &$\left(\begin{array}{ccc}0.0&-1.5&-3.2\\-1.5&-18.7&-35.1\\-3.2&-35.1&7.2\end{array}\right)$&$\left[\begin{array}{c}-43.3\\31.7\\0.1\end{array}\right]$&$\left[\begin{array}{c}(-0.07,-0.82,-0.57)\\(-0.06,-0.57,0.82)\\(1.00,-0.09,0.01)\end{array}\right]$&$\left[\begin{array}{c}16840\\16915\\16884\end{array}\right]$&$\left[\begin{array}{c}15729\\15804\\15773\end{array}\right]$\\
$0^+$ &$\left(\begin{array}{cc}-16.5&60.7\\60.7&24.8\end{array}\right)$&$\left[\begin{array}{c}68.3\\-60.0\end{array}\right]$&$\left[\begin{array}{c}(0.58,0.81)\\(-0.81,0.58)\end{array}\right]$&$\left[\begin{array}{c}16952\\16823\end{array}\right]$&$\left[\begin{array}{c}15841\\15713\end{array}\right]$\\
\hline
\end{tabular}
\end{table}

We show the results for the $bb\bar{b}\bar{c}$ (and $cb\bar{b}\bar{b}$) states in Table \ref{result:bbbbarcbar} and the spectrum for $bb\bar{b}\bar{c}$ in Fig. \ref{fig} (d). The maximum mass splitting is around 130 MeV. Comparing with the former systems, the Pauli principle works only for one diquark now, which leads to one more $1^+$ tetraquark.

From Fig. \ref{fig} (d), we may easily understand the rearrangement decay behaviors for the $bb\bar{b}\bar{c}$ states. The two scalar states have similar properties to the states in the former systems. For the $J=2$ tetraquark, there is one more $D$-wave decay channel compared to the $cc\bar{b}\bar{b}$ case because the violation of the heavy quark spin symmetry results in the non-degeneracy for the thresholds of $\eta_bB_c^*$ and $\Upsilon B_c$. The interesting observation appears for the $J^P=1^+$ states. The color-spin mixing affects the masses of the states relatively largely. The resulting observation is: the highest and the intermediate states are kinematically allowed to decay into $\Upsilon B_c$ and $\eta_bB_c^*$ channels, while the lowest state has no rearrangement decay channel. From the relative positions for the solid/black, dashdotted/blue, and dashed/red lines, the interactions within the diquarks are effectively attractive for the lowest $1^+$ tetraquark, repulsive for the intermediate $1^+$ state and also attractive for the highest $1^+$ state. This feature is a result of balance between attraction/repulsion in $bc$, repulsion in $bb$, attraction between quarks and antiquarks, and channel coupling. From the effective interaction within the diquarks and the estimated mass, the lowest $1^+$ state is a good candidate of stable tetraquarks. Apparently, this state looks like an excited $B_c$ where a $b\bar{b}$ pair is excited (see discussions in Ref. \cite{Chen:2016ont} for other tetraquarks). By checking the CMI matrix elements, one finds that the stable state is possible mainly because of the attraction within the $bc$ diquark. The situation is similar to the $T_{cc}$ state ($cc\bar{q}\bar{q}$) where the attractive $ud$ diquark contributes dominantly \cite{Hyodo:2012pm,Hyodo:2017hue}.

With the observed bottomonium and $B_c$ states, in principle, resonance structures above the $\eta_b B_c$ threshold can all be investigated. Once experiments collect enough $B_c$ data, interesting states in the $\Upsilon B_c$ channel would probably be observed first.

\subsection{The $cc\bar{c}\bar{b}$ and $bc\bar{c}\bar{c}$ systems}

We may call such tetraquarks as ``mirror'' partners of the previous states. The color-spin structure is the same as the $bb\bar{b}\bar{c}$
system but the decay feature relies on masses and may be different. We present the numerical results in Table \ref{result:cccbarbbar} and draw the spectrum in Fig. \ref{fig} (e). The maximum mass splitting between the two scalar tetraquarks is around 180 MeV. Rearrangement decays for these tetraquarks are possible only when the mass is high enough.

The rearrangement decay properties for the $J=2$ and $J=0$ tetraqaurks are similar to those for the $bb\bar{b}\bar{c}$ states.
For the three axial vector tetraquarks, the low mass one has only one decay channel $\eta_cB_c^*$, the intermediate one has two
$\eta_cB_c^*$ and $\psi B_c$, and the high mass state has one more channel $\psi B_c^*$. Contrary to the $bb\bar{b}\bar{c}$ case, the effective interaction within the diquarks in the lowest $1^+$ state is repulsive, which is a result that the $cc$ interaction is stronger than the $bb$ interaction. If this state does exist, it should be less stable than the lowest $1^+$ $bb\bar{b}\bar{c}$.

Early experimental investigations on possible resonances in the $cc\bar{c}\bar{b}$ system should be through the channel $\psi B_c$, which means that four tetraquarks, a high mass scalar, two axial vectors, and one tensor, could be observed first.

\begin{table}[h!]
\caption{Results for the $cc\bar{c}\bar{b}$ and $bc\bar{c}\bar{c}$
systems in units of MeV with the approximation $C_{QQ}=C_{Q\bar{Q}}$. The masses in the fifth column are
estimated with $m_b=5052.9$ MeV and $m_c=1724.8$ MeV. The last column
lists masses estimated from the $(\psi B_c)$ threshold. The base for
the  $J=1$ case is $(\phi_2\chi_2, \phi_2\chi_4, \phi_1\chi_5)^T$
and that for the $J=0$ case is
$(\phi_{2}\chi_{3},\phi_{1}\chi_{6})^T$.}\label{result:cccbarbbar}
\centering
\begin{tabular}{c|ccccc}\hline
$J^P$ & $\langle H_{CM} \rangle$ &Eigenvalue &Eigenvector &Mass&$(\psi B_c)$\\
$2^+$ &45.9&45.9&1&10273&9442\\
$1^+$ &$\left(\begin{array}{ccc}0.0&7.5&16.0\\7.5&-12.3&-48.6\\16.0&-48.6&16.8\end{array}\right)$&$\left[\begin{array}{c}54.4\\-53.1\\3.3\end{array}\right]$&$\left[\begin{array}{c}(0.16,-0.57,0.81)\\(0.28,-0.76,-0.59)\\(0.95,0.32,0.04)\end{array}\right]$&$\left[\begin{array}{c}10282\\10174\\10231\end{array}\right]$&$\left[\begin{array}{c}9451\\9343\\9400\end{array}\right]$\\
$0^+$ &$\left(\begin{array}{cc}-22.9&84.3\\84.3&34.4\end{array}\right)$&$\left[\begin{array}{c}94.7\\-83.3\end{array}\right]$&$\left[\begin{array}{c}(0.58,0.81)\\(-0.81,0.58)\end{array}\right]$&$\left[\begin{array}{c}10322\\10144\end{array}\right]$&$\left[\begin{array}{c}9491\\9313\end{array}\right]$\\
\hline
\end{tabular}
\end{table}

\subsection{The $bc\bar{b}\bar{c}$ system}

We show the results for the twelve states in Table \ref{result:bcbbarcbar}. Now the maximum mass splitting (240 MeV) again occurs between the scalar  tetraquarks. In estimating the tetraquark masses, there are two reference thresholds we can use, $\Upsilon\psi$ and $B_c^+B_c^-$. As the investigations in other systems \cite{Chen:2016ont,Wu:2017weo}, the former threshold leads to lighter masses that can be treated as a lower limit on the theoretical side. We here adopt the masses estimated with the latter threshold. Future measurements may answer whether the scheme is reasonable or not. Seven meson-meson channels are involved in discussing decay properties with the obtained results. We show the spectrum and these channels in Fig. \ref{fig} (f).

\begin{table}[htbp]
\caption{Results for the $bc\bar{b}\bar{c}$ system in units of MeV with the approximation $C_{QQ}=C_{Q\bar{Q}}$.The masses in the fifth column are
estimated with $m_b=5052.9$ MeV and $m_c=1724.8$ MeV.  The last two columns list masses estimated from the $(\Upsilon\psi)$ and the $(B_cB_c)$ thresholds. The bases for the
$J=2$ and $J=0$ cases are $(\phi_1\chi_1,\phi_2\chi_1)^T$ and $(\phi_1\chi_3,\phi_1\chi_6,\phi_2\chi_3,\phi_2\chi_6)^T$, respectively. If $J=1$, the bases for the cases $C=+$ and $C=-$ are $({[\phi\chi]}_+^{6\bar{6}},{[\phi\chi]}_+^{\bar{3}3})^T$ and $(\phi_1\chi_2, \phi_2\chi_2,
{[\phi\chi]}_-^{6\bar{6}},{[\phi\chi]}_-^{\bar{3}3})^T$, respectively.}\label{result:bcbbarcbar}
\centering
\begin{tabular}{c|cccccc}\hline
$J^{PC}$ & $\langle H_{CM} \rangle$ &Eigenvalue &Eigenvector &Mass&$(\Upsilon\psi)$& $(B_cB_c)$\\
$2^{++}$ &$\left(\begin{array}{cc}40.5&4.5\\4.5&37.3\end{array}\right)$&$\left(\begin{array}{c}43.7\\34.1\end{array}\right)$&$\left[\begin{array}{c}(-0.82,-0.58)\\(0.58,-0.82)\end{array}\right]$&$\left(\begin{array}{c}13599\\13590\end{array}\right)$&$\left(\begin{array}{c}12557\\12548\end{array}\right)$&$\left(\begin{array}{c}12700\\12690\end{array}\right)$\\
$0^{++}$ &$\left(\begin{array}{cccc}-107.5&-9.2&9.1&72.5\\-9.2&26.4&72.5&0.0\\9.1&72.5&-21.9&-3.7\\72.5&0.0&-3.7&-52.8\end{array}\right)$&$\left(\begin{array}{c}-159.4\\78.8\\-72.7\\-2.4\end{array}\right)$&$\left[\begin{array}{c}(0.82,0.08,-0.11,-0.56)\\(-0.02,0.81,0.58,-0.03)\\(0.08,-0.58,0.80,-0.14)\\(0.57,-0.01,0.08,0.82)\end{array}\right]$&$\left(\begin{array}{c}13396\\13634\\13483\\13553\end{array}\right)$&$\left(\begin{array}{c}12354\\12592\\12441\\12511\end{array}\right)$&$\left(\begin{array}{c}12496\\12735\\12583\\12653\end{array}\right)$\\
$1^{++}$ &$\left(\begin{array}{cc}14.1&-41.9\\-41.9&-15.5\end{array}\right)$&$\left(\begin{array}{c}-45.1\\43.7\end{array}\right)$&$\left[\begin{array}{c}(0.58,0.82)\\(-0.82,0.58)\end{array}\right]$&$\left(\begin{array}{c}13510\\13599\end{array}\right)$&$\left(\begin{array}{c}12468\\12557\end{array}\right)$&$\left(\begin{array}{c}12611\\12700\end{array}\right)$\\
$1^{+-}$ &$\left(\begin{array}{cccc}-58.1&4.5&-16.0&13.6\\4.5&-2.1&13.6&-6.4\\-16.0&13.6&3.5&41.9\\13.6&-6.4&41.9&-19.7\end{array}\right)$&$\left(\begin{array}{c}-77.7\\36.8\\-35.5\\-0.1\end{array}\right)$&$\left[\begin{array}{c}(-0.73,0.16,-0.43,0.50)\\(0.05,-0.18,-0.80,-0.56)\\(0.68,0.19,-0.41,0.58)\\(0.00,0.95,0.00,-0.31)\end{array}\right]$&$\left(\begin{array}{c}13478\\13592\\13520\\13555\end{array}\right)$&$\left(\begin{array}{c}12436\\12550\\12478\\12513\end{array}\right)$&$\left(\begin{array}{c}12578\\12693\\12620\\12656\end{array}\right)$\\
\hline
\end{tabular}
\end{table}

Since there is no constraint from the Pauli principle, the obtained spectrum is more complicated than other systems. All the masses are above the lowest meson-meson threshold and they should decay. At least four rearrangement decay channels for the $2^{++}$ states are $D$-wave channels. If the $2^{++}$ state is above the $B_c^{*+}B_c^{*-}$ threshold, the $S$-wave channel is also opened. For the channels with $J^{PC}=1^{++}$, the $S$-wave $\Upsilon\psi$ is the lowest one. The decays of the two $1^{++}$ tetraquarks into this channel are both allowed. For the four $1^{+-}$ states, they all have $S$-wave decay channels $\Upsilon\eta_c$ and $\eta_b\psi$. There is at least one allowed rearrangement decay channel, $\eta_b\eta_c$, for the four $0^{++}$ tetraquarks through $S$-wave interaction. It seems that all the $bc\bar{b}\bar{c}$ tetraquarks are not stable.

On the other hand, one finds that the interactions within the diquarks in several states are effectively attractive. This indicates that in the competition between the interactions of $QQ$ and $Q\bar{Q}$, the stronger attraction in $QQ$ may make such states relatively stable. The lowest $0^{++}$ state satisfies this condition and it has only one rearrangement decay channel. Maybe this state has relatively narrow width. The lower $1^{++}$ state has also similar features. Its $S$-wave decay channel is $\Upsilon\psi$ and this tetraquark is worthwhile study. Because of the existence of possible tetraquark states, searching for exotic phenomena with $\Upsilon\psi$ may help us to understand the strong interactions between heavy quarks. In Ref. \cite{Richard:2017vry}, Richard et al also observe that bound $bc\bar{b}\bar{c}$ might be more favorable than $bb\bar{b}\bar{b}$ and $cc\bar{c}\bar{c}$.

\section{Discussions and summary}\label{sec4}

In the chiral quark model, the interaction between the light quarks may also arise from the exchange of Goldstone bosons. For the pure heavy systems, such an interaction is absent. Therefore, one needs to consider only the gluon-exchange interaction for the present $Q_1Q_2\bar{Q}_3\bar{Q}_4$ systems. As a short range force, it helps to form compact tetraquarks rather than meson-meson molecules if bound four-quark states do exist. We have studied the mass splittings of these tetraquark states with the simple color-magnetic interaction in this work.

For the case ($bc\bar{b}\bar{c}$ system) without constraint from the Pauli principle, the number of color-singlet tetraquarks with $6_c$ diquark is equal to that with $\bar{3}_c$ diquark. The two color structures may couple through the color-magnetic interaction. From Table \ref{result:bcbbarcbar}, the coupling for the $2^{++}$ states is weak while that for other states is stronger. For the case with constraint from the Pauli principle, the number of color-singlet tetraquarks with the $\bar{3}_c$ diquark is bigger than that with $6_c$ diquark. Their mixing is generally significant (see Tables \ref{result:bbbb}-\ref{result:cccbarbbar}). In both cases, the tetraquarks with $6_c$ diquark do not exist independently.

After the configuration mixing effects are considered, both the heaviest tetraquark and the lightest tetraquark for a system are the scalar states. The maximum mass differences are around 120 MeV, 220 MeV, 140 MeV, 130 MeV, 180 MeV, and 240 MeV for the $bb\bar{b}\bar{c}$, $cc\bar{c}\bar{c}$, $bb\bar{c}\bar{c}$, $bb\bar{b}\bar{c}$, $cc\bar{c}\bar{b}$, and $bc\bar{b}\bar{c}$ systems, respectively. Other states fall in these mass difference ranges. Whether the
tetraquarks decay through quark rearrangements and how many thresholds fall in these ranges rely on the tetraquark masses. For comparison, we estimate the masses with two approaches. We mainly discuss the tetraquark properties by using the masses estimated with the threshold of some meson-meson state, the second approach.

Since the parameter $C_{QQ}$ cannot be extracted with experimental data, we use two approximations to determine them. We also show results in the extreme case $C_{QQ}=0$. From the comparison study, we find that (1) the estimated tetraquark masses are affected slightly and (2) the plotted spectra can be easily used to judge which states contain effectively attractive diquarks. From the viewpoint that narrow compact tetraquarks should have attractive diquarks inside them and have as few $S$-wave decay channels as possible, we find that most $QQ\bar{Q}\bar{Q}$ states are not stable. However, a stable $1^+$ $bb\bar{b}\bar{c}$ state is observed and relatively narrow $bc\bar{b}\bar{c}$ tetraquarks are also possible in the present model investigations. The latter system is also proposed to have bound states in Ref. \cite{Richard:2017vry}. Although the model we use is simple and it does not involve dynamics, the basic features of the obtained spectra might be somehow reasonable. To give more reliable results, the present model needs to be improved.

If experiments could observe one resonant state in the channel of two heavy quarkonia, its nature as a tetraquark is favored. More tetraquarks should also exist and searches for them are strongly called for. We hope the decay channels discussed in this paper are helpful for the experimental search.

To summarize, we have explored the mass splittings between the $Q_1Q_2\bar{Q}_3\bar{Q}_4$ tetraquarks with a simple model. The mixing between different color structures are considered. We have estimated their masses with two approaches and discussed possible rearrangement decay channels. Stable or narrow tetraquarks in the $bb\bar{b}\bar{c}$ and $bc\bar{b}\bar{c}$ systems are worthwhile study in heavy meson-meson channels such as $\Upsilon\psi$. Hopefully, the exotic tetraquark states composed of four heavy quarks may be observed at LHCb in the future.

\section*{Acknowledgments}
YRL thanks F. Huang (UCAS, Beijing) for a discussion and people at Tokyo Institute of Technology for their hospitality where the updated draft was finalized. This project is supported by National Natural Science Foundation of
China under Grants No. 11775132, No. 11222547, No. 11175073, No.
11261130311 and 973 program. XL is also supported by the National Program for Support of Top-notch Young Professionals.

\end{document}